**Electronic mobility, doping, and defects in epitaxial BaZrS$_3$ chalcogenide perovskite thin films**

Jack Van Sambeek[1], Jessica Dong[1], Anton V. Ievlev[2], Tao Cai[1], Ida Sadeghi[1], R. Jaramillo[1]

1. Department of Materials Science and Engineering, Massachusetts Institute of Technology

2. Center for Nanophase Material Sciences, Oak Ridge National Laboratory, Oak Ridge USA

**Abstract**

We present the electronic transport properties of BaZrS$_3$ (BZS) thin films grown epitaxially by gas-source molecular beam epitaxy (MBE). We observe *n*-type behavior in all samples, with carrier concentration ranging from $4 \times 10^{18}$ to $4 \times 10^{20}$ cm$^{-3}$ at room temperature (RT). We observe a champion RT Hall mobility of 11.1 cm$^2$V$^{-1}$s$^{-1}$, which is competitive with established thin-film photovoltaic (PV) absorbers. Temperature-dependent Hall mobility data show that phonon scattering dominates at room temperature, in agreement with computational predictions. X-ray diffraction data illustrate a correlation between mobility and stacking fault concentration, illustrating how microstructure can affect transport. Despite the well-established environmental stability of chalcogenide perovskites, we observe significant changes to electronic properties as a function of storage time in ambient conditions. With the help of secondary-ion mass-spectrometry (SIMS) measurements, we propose and support a defect mechanism that explains this behavior: as-grown films have a high concentration of sulfur vacancies that are shallow donors ($V_S^{\cdot}$ or $V_S^{\cdot\cdot}$), which are converted into neutral oxygen defects ($O_S^{\times}$) upon air exposure. We discuss the relevance of this defect mechanism within the larger context of chalcogenide perovskite research, and we identify means to stabilize the electronic properties.

**Introduction**

Chalcogenide perovskite semiconductors are of interest as absorbers for thin-film photovoltaics (PV). Chalcogenide perovskites are distinguished by their Earth-abundant elements, low-to-no toxicity, thermo-chemical stability, strong optical absorption, and direct band gap ($E_g$) that can be tuned continuously by alloying.[1–7] BaZrS$_3$ (BZS, with $E_g$ = 1.9 eV) is by far the most-studied chalcogenide perovskite, with decades of publications reporting synthesis of samples with different form factors (powder, single crystals, nanoparticles, polycrystalline thin films, epitaxial thin films) and properties including photoluminescence and dielectric susceptibility[1,3,5,7–20]. However, the electronic properties of BZS remain little-reported. Measuring and understanding the electronic properties of BZS, and by extension of chalcogenide perovskites more broadly, is an important next step in the development of these materials for PV.

Published reports on the electronic properties of chalcogenide perovskites suggest that they are likely but not necessarily *n*-type. A recent computational study by Yuan *et al.* predicts that the dominant intrinsic dopants in BZS are S vacancies, which are shallow double donors and are responsible for *n*-type conductivity. Yuan *et al.* also predict that the electron mobility ($\mu$) at room temperature (RT) is phonon-limited to 37 cm$^2$V$^{-1}$s$^{-1}$.[21] These theoretical predictions are consistent with experimental observations: at least two publications report large electron concentration ($n > 10^{20}$ cm$^{-3}$) in BZS thin films, and Aggarwal *et al.* provide evidence that *n* is positively correlated with S deficiency.[12,22] Hanzawa *et al.* reported synthesis of both *n*- and *p*-type SrHfS$_3$ by aliovalent doping.[23] This is encouraging for achieving ambipolar doping in chalcogenide perovskites more broadly, especially because SrHfS$_3$ has a rather wide band gap (2.3 eV), that ought to make it more difficult to achieve ambipolar doping than for compounds with smaller band gap. Less is known about the mobility and how it might be improved in thin films. *n*-type mobility exceeding 10 cm$^2$V$^{-1}$s$^{-1}$ has been reported.[12] Experimental studies of the mobility-limiting mechanisms have not been reported.

## Results

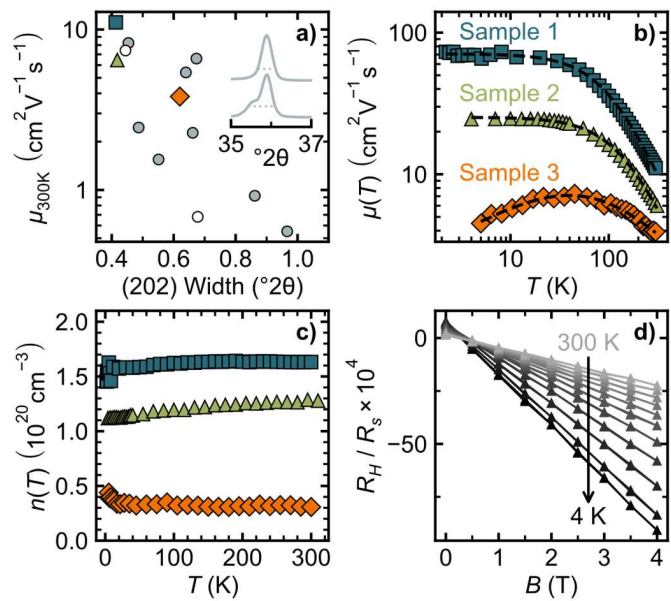

**Figure 1: Hall effect measurement results for a distribution of BaZrS$_3$ thin films. (a)** Hall mobility at 300 K plotted against HRXRD-measured (202) full-width at one quarter max (FWQM). The inset shows the (202) peak shape for the two samples indicated by hollow markers, with horizontal dashed lines showing their FWQM. **(b)** Temperature-dependent Hall mobility for Samples 1-3. Dashed curves show fits to the expression: $\mu^{-1} = AT^{\alpha} + BT^{\beta}$, which is a Matthiessen's rule addition of two power-law mobility trends. For Samples 1 and 2, $\alpha$ is -1.6 and $\beta$ is 0. For Sample 3, $\alpha$ and $\beta$ are -0.7 and 0.4, respectively. **(c)** Temperature-dependent electron concentration for Samples 1-3. **(d)** Field-dependence of Hall measurements for Sample 2 shown at temperatures in the range 4 K to 300 K. The ordinate is Hall resistance ($R_H$) divided by sheet resistance ($R_S$), scaled by a factor of $10^4$. In all panels, squares correspond to Sample 1, triangles to Sample 2, and rotated squares to Sample 3. Circles indicate other samples for which further characterization is not included in this report.

In this work, we leverage a large set of epitaxial BZS thin films, and complementary experiments including Hall effect measurements, X-ray diffraction, and secondary ion mass spectroscopy (SIMS), to gain a more complete understanding of electronic transport. As grown, our BZS films have widely-varying electronic properties, with four-point probe measured sheet resistance ($R_S$) values ranging from below 1 kΩ/□, to more than 1 GΩ/□. In **Fig. 1** we present Hall measurement results for a large number of samples that are sufficiently conductive ($R_S < \sim$1 MΩ/□) for our instrumentation. These data were recorded shortly after synthesis of each sample. We observe exclusively *n*-type behavior, with large RT carrier concentration $n$(300 K) in the range $4 \times 10^{18}$ - $4 \times 10^{20}$ cm$^{-3}$ (**Fig. S1**). We observe a champion RT Hall mobility $\mu$(300 K) of 11.1 cm$^2$V$^{-1}$s$^{-1}$, which is on par with established thin-film PV absorbers including lead-halide perovskites.[24] This champion film (Sample 1) also shows strong room-temperature PL, suggesting a relatively slow rate of non-radiative recombination (**Fig. S2**).

We observe a wide distribution in $\mu$ across all films measured, extending to below 0.5 cm$^2$V$^{-1}$s$^{-1}$. In general, we expect that $\mu$ will be lower for films with broader x-ray diffraction peak widths, as peak width grows with increasing concentrations of extended defects. Several of the lesser-quality films show a low-angle shoulder on the (202) XRD peak (**Fig. 1a**, inset). Our previous reports attributed this (202) peak shoulder to the presence of antiphase boundary (APB) defects, which are expected to impede carrier transport. To illustrate the correlation of transport with film defect concentration, in **Fig. 1a** we plot $\mu$ at 300 K against the full-width at one quarter max (FWQM) of the (202) peak. FWQM is sensitive to symmetric peak broadening and to the low-angle shoulder, unlike the more commonly-used full-width at half-maximum (FWHM) that does not well capture low-intensity shoulders. The data in **Fig. 1a** show that $\mu$ is strongly reduced with increasing FWQM. **Fig. 1a** highlights the need for continuous improvement in film growth process control.

We performed temperature-dependent Hall measurements on a subset of the films shown in **Fig. 1a** to investigate the dominant mobility-limiting mechanisms. In **Fig. 1b** we present temperature-dependent Hall mobility data ($\mu(T)$) for three samples, referred henceforth as Samples 1–3. We find that the electron concentration for each sample ($n(T)$) is approximately temperature-independent across the entire measured range, from 4 – 300 K (**Fig. 1c**). Such weak temperature dependence of $n(T)$ is expected in the degenerate doping limit, and has been reported previously in this material system[22].

Near room temperature, all three samples show decreasing $\mu$ with increasing temperature, which identifies phonon scattering as the mobility-limiting mechanism, consistent with recent computational predictions[21,27]. Samples 1 and 2, with larger carrier concentration and lower (202) peak FWQM, display prototypical features of metallic conduction: phonon-limited mobility near room temperature, transitioning to temperature-independent, defect-limited mobility at cryogenic temperature. Sample 3, with smaller carrier concentration and larger (202) peak FWWM, shows $\mu(T)$ behavior intermediate between the metallic and non-degenerate doping limits, possibly due to its lower carrier density. Upon cooling, $\mu(T)$ reaches a maximum value of 7 cm$^2$V$^{-1}$s$^{-1}$ around 50 K. Further cooling causes $\mu$ to decrease, though at a slower rate than the $T^{3/2}$ behavior expected from ionized-impurity scattering in non-degenerate semiconductors.[27] In contrast to these three epitaxial samples, we observe non-metallic behavior in both $n$ and $\mu$ for a polycrystalline film sample (Sample P1) grown on Al$_2$O$_3$, despite its nearly 10 times larger room-temperature $n$ of $3 \times 10^{21}$ cm$^{-3}$ (**Fig S3**). Sample P1 shows strongly temperature-dependent $n$ over a two-decade range, and $\mu(T)$ varies between 0.1 to 0.9 cm$^2$V$^{-1}$s$^{-1}$ with low-temperature behavior indicative of ionized-impurity scattering. This unexpected result suggests that perhaps high-angle and random grain boundaries change the conditions for metallicity in BaZrS$_3$.

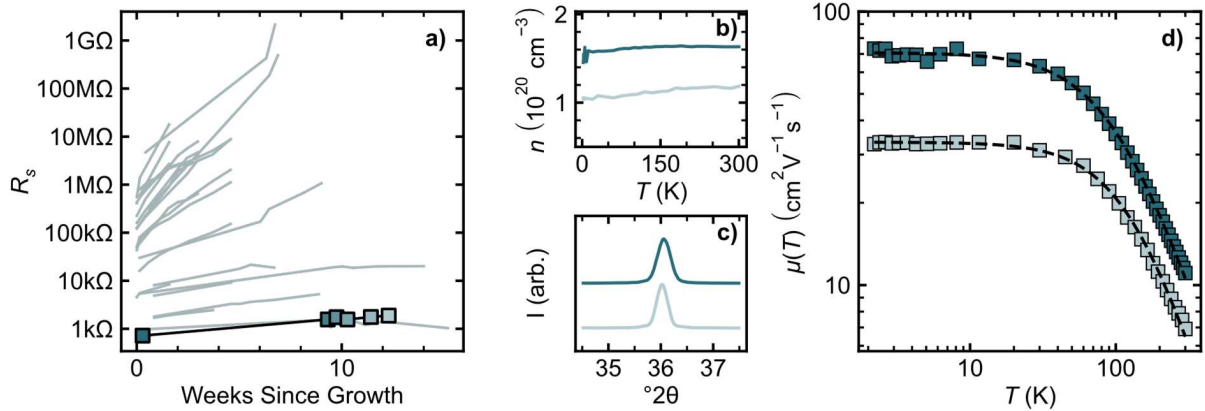

**Figure 2: Time-evolution of properties after storage in ambient conditions. (a)** Repeated 4-point probe $R_s$ measurements for Sample 1 (blue squares) and a large set of BZS films (light grey lines) as a function of time from sample fabrication. **(b)** Temperature-dependent electron concentration, as-grown and after 75 days of storage for Sample 1. **(c)** HRXRD measurements as-grown and after 285 days of storage for Sample 1. **(d)** Temperature-dependent Hall mobility, as-grown and after 75 days of storage for Sample 1. Dashed curves show fits to the same dual-power law expression used in Fig. 1b; here $\alpha$ is -1.6 as-grown and -1.8 after storage, with $\beta$ being 0 in both cases. In panels **b-d**, the upper, darker data are the as-grown measurement; the lighter, lower data are the measurement after storage.

Despite the established air-stability of BaZrS$_3$, and of chalcogenide perovskites broadly, we observe significant instabilities in electrical properties during storage under ambient conditions. $R_S$ for each sample increases monotonically over time, with a rate that varies strongly sample-to-sample. For some samples, $R_S$ increases as quickly as one decade per week, and for others as slowly as doubling over several months (**Fig. 2a**). This behavior occurs without any associated structural changes. In **Fig. 2** we demonstrate the effects of storage in ambient conditions on Sample 1, as a representative case. For this sample, the four-point-probe-measured $R_S$ increased from 720 Ω/□ as-grown, to 1.6 kΩ/□ after 75 days of storage. We recorded HRXRD measurements of the film (202) peak within one day of growth, and repeated 285 days later. We observed no significant change in peak shape or position (**Fig. 2c**), demonstrating little-to-no structure change due to long-term storage in air. To understand the observed changes in electronic transport, we performed temperature-dependent Hall measurements on this sample, once three days after synthesis (data in **Fig. 1**), and again 72 days later. We compare the $n(T)$ curves in **Fig 2b**. In both measurements, we observe $n$-type behavior with $n$ nearly independent of temperature. We observe a 25% reduction in $n$ after sample storage, from $1.6 \times 10^{20}$ cm$^{-3}$ to $1.2 \times 10^{20}$ cm$^{-3}$. $\mu(T)$ data (**Fig. 2d**) shows that $\mu$ is lower after storage but also that the temperature-dependence $\mu(T)$ is unchanged, suggesting that the dominant mobility-limiting mechanisms remained the same.

To explain these effects of sample storage in air, we hypothesize a simple defect mechanism,

$$V_S^{\cdot\cdot} + 2e^- + \frac{1}{2}O_{2_{(g)}} \rightarrow O_S^\times \tag{1}$$

in which ambient oxygen fills vacant anion sites, converting $V_S$ donors to $O_S$ defects, which are predicted to be electrically neutral.[21] Because the films are very thin (< 50 nm), and the sulfur vacancy concentration is quite large (approaching 1% of sulfur sites), it is reasonable to expect rapid anion diffusion, which explains how oxygen exposure at room temperature can affect film properties on the time scale of weeks.

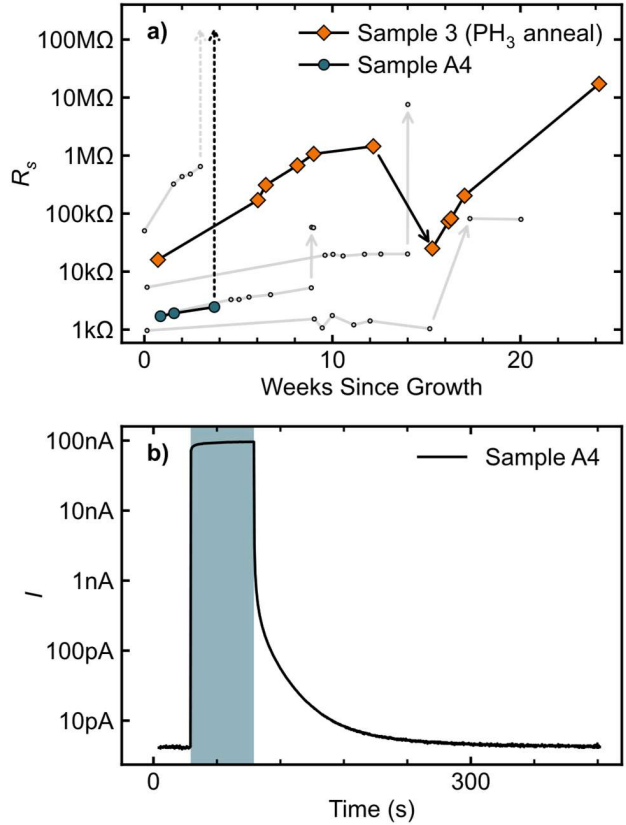

**Figure 3: Effects of annealing on electronic properties. a)** Repeated four-point probe $R_s$ measurements as a function of time from fabrication for Samples A1–A5 and Sample 3. Solid connecting lines shows changes during storage in ambient conditions, while arrows show changes brought on by annealing treatments. Light grey data is for Samples A1–A3 and A5. Samples A3 and A4 were immeasurably resistive after annealing, indicated by the broken arrows. **b)** Post-anneal photocurrent measurement of Sample A4 under 1 V of applied bias. The sample was illuminated by a white LED light source with an irradiance of 100 mW·cm$^{-2}$ during the period indicated by the shaded rectangle.

We conducted annealing experiments to further understand the aging mechanisms and to explore post-growth defect control. Samples A1 – A4 were annealed in a vacuum tube furnace under a range of H$_2$S gas flow (0 to 10 sccm) and a range of temperature (150 to 500 °C). Sample A5 was annealed in an ultra-high vacuum chamber at 200 °C while exposed to an H$_2$S plasma. We report detailed annealing conditions in **Table S1**. All annealed samples showed at least one order-of-magnitude increase in $R_S$ following annealing, with Samples A3 and A4 becoming immeasurably resistive (**Fig. 3a**). Only Samples A2 and A5 remained conductive enough to enable post-anneal Hall measurements. These samples showed reductions in both $n$ (1.2 × 10$^{20}$ cm$^{-3}$ to 1.9 × 10$^{19}$ cm$^{-3}$ for Sample A2, and 6.9 × 10$^{20}$ cm$^{-3}$ to 6.7 × 10$^{19}$ cm$^{-3}$ for Sample A5) and $\mu$ (8.2 cm$^2$V$^{-1}$s$^{-1}$ to 1.1 cm$^2$V$^{-1}$s$^{-1}$ for Sample A2, and 2.3 cm$^2$V$^{-1}$s$^{-1}$ to 0.32 cm$^2$V$^{-1}$s$^{-1}$ for Sample A5). The reduction in $\mu$ is notable as this would be unexpected if the annealing process was incorporating sulfur into the film to heal V$_S$ defects. Instead, it seems likely that the annealing treatments are simply accelerating the aging effects observed during ambient storage by converting V$_S$ to O$_S$. This may be possible even in oxygen-free annealing environments if the film native oxide layer (see below) acts as an oxygen source during annealing.

In addition to Samples A1-A5, Sample 3 was annealed at 400 °C in a load-locked semiconductor processing furnace under a high flow rate of $PH_3$ gas. This processing scheme should limit oxygen reactivity in comparison to the other processing methods, owing to the superior furnace quality and more reducing atmosphere. Indeed, we observe that unlike the other annealed samples, Sample 3's resistance was significantly lowered by annealing (**Fig. 3a**), akin to the annealing outcomes recently reported by Aggarwal *et al.*[22] We suspect that this annealing treatment had minimal $V_S$ to $O_S$ conversion, and instead increased the $V_S$ concentration either by liberating oxygen from existing $O_S$ defects or by removing sulfur from the film. After annealing, $R_S$ once again began rapidly increasing over time while the sample was stored in ambient conditions, consistent with our proposed $V_S$ to $O_S$ defect mechanism.

Post-anneal Hall measurements were not possible on Samples A1, A3, and A4, so changes to $n$ and $\mu$ for these samples cannot be directly quantified. In **Fig. 3b** we show the post-anneal photoconductivity response for Sample A4, which had $R_S$ = 1.7 kΩ/□ as-grown, and became immeasurably resistive after annealing. The four-decade-large photocurrent response measured post-anneal shows that the sample still has finite mobility and useful semiconductivity, confirming that the immeasurably large $R_S$ is due to a reduction in the $\mu n$ product, rather than morphological changes disrupting the film's connectivity. This finding agrees with scanning electron microscopy (SEM) micrographs of as-grown and post-anneal films, showing no appreciable microstructural changes (**Fig. S6**). However, without specific knowledge of changes in $n$ and $\mu$, the photoconductivity measurements alone are inconclusive as to whether the high-resistance samples underwent the same oxygen incorporation mechanism as suspected for Samples A2 and A5. We must turn to compositional measurements to further understand the defect mechanisms at play.

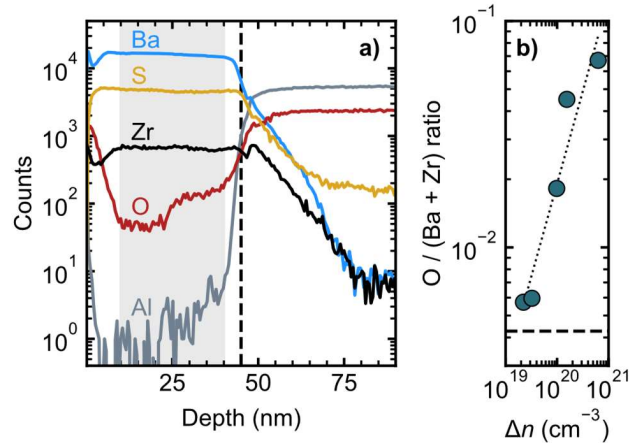

**Figure 4: Identifying the effect of oxygen uptake on carrier concentration. (a)** ToF-SIMS depth profiles for Sample A1 showing count rates for Ba, S, Zr, O, and Al secondary ions as a function of etching depth. The vertical dashed line indicates the film-substrate interface. The shaded grey area marks the depth range used for averaging. **(b)** Trend of oxygen content *vs.* reduction in carrier concentration due to annealing. The ordinate shows the ratio of O signal to combined Ba and Zr signal measured by ToF-SIMS. The dashed horizontal line shows the same average for an unannealed film measured 85 days after it was grown. The abscissa shows the reduction in electron concentration observed between pre- and post-anneal measurements. The dashed line is a power-law fit with an exponent of 0.8 (0.1 standard error).

To augment the electronic measurements of annealed samples, we performed time-of-flight secondary ion mass spectrometry (ToF-SIMS) depth profiling of Samples A1–A5, and an additional sample that was not annealed (Sample N). Due to a lack of suitable reference films of standard composition, the ToF-SIMS results are semiquantitative. While measured secondary ion intensities are linear to actual element concentrations, it impossible to calculate actual concentrations from ToF-SIMS data only, without calibration. Therefore, we cannot quantify atomic concentrations, but can analyze relative composition comparisons between samples. We calibrated ion etch rates using atomic-force microscopy on Sample N (**Fig. S5**) and used the measured etch rate to determine the measurement depth for each cycle in the ToF-SIMS measurements. Film thicknesses were determined from the depth-calibrated SIMS data and are shown in **Table S1**. We present in **Fig. 4a** depth-resolved elemental signal intensity line scans for Sample A1 (see **Fig. S4** for line scans of all other samples). An oxygen-rich layer is apparent near the film surface, with rapidly decreasing oxygen signal as depth increases. This layer is consistent with our prior report of a self-limiting native oxide.[13] Beyond this surface layer, the signals of Ba, Zr, and S are well defined and constant until reaching the substrate. Oxygen is clearly present in the bulk of the film with non-uniform signal intensity. Analysis of Samples A2, A4, and A5 is complicated by the incomplete film coverage of these samples (**Fig. S7**). For these partially-open films, O signal observed within the depth of the film could either be from impurities in the film, or from exposed regions of the oxide substrate. We discuss this complication in Supplementary Note 1; we conclude that excluding the bottom 5 nm of each film allows us to confidently attribute the measured O signal to the film rather than the substrate.

We use the ToF-SIMS data to test the hypothesis that annealing-induced electronic changes are due to O incorporation. If annealing converts $V_S$ to $O_S$ defects, then the observed decrease in $n$ should correlate with the post-anneal O concentration. The log-log plot presented in **Fig. 4b** shows the expected trend, with a power law fit yielding a near-linear exponent of $0.8 \pm 0.1$. The ordinate is the average ratio between oxygen and combined Ba and Zr signals in the bulk of each film, which we use as a proxy for the ratio between oxygen concentration and anion site density. The abscissa is the $n$ reduction as a result of annealing, according to Hall measurements. For films that were too resistive for post-anneal Hall measurements, we assume that post-anneal $n$ was much smaller than the initial $n$. We list in **Table S1** the values of pre- and post-anneal $n$ for each sample. The clear correlation shown in **Fig. 4b** is consistent with the proposed defect mechanism.

**Discussion and Conclusion**

Our results emphasize the importance of continued improvements in film growth process control. We and others have demonstrated control over chalcogenide perovskite film phase, but not yet over defects. We attribute the large sample-to-sample variation in electronic properties to variations in intrinsic defect concentrations, including but not necessarily limited to sulfur vacancies. Sulfur interstitials may also be undesirable non-radiative recombination centers.[21] There are several promising approaches to controlling sulfur defect concentrations, including post-growth annealing. We recently found that annealing in $H_2S$ plasma immediately following film growth, before air exposure, may yield films with orders-of-magnitude smaller carrier concentration and superior air stability (unpublished results). Reducing film growth temperature is also likely to lower the vacancy concentration. There is also likely need to improve control over cation defects in this ternary system, although the impacts of Ba- and Zr-based defects on film properties remain uncertain. We expect that continued innovations in chemical synthesis pathways, coupled with more robust processing controls, will produce substantial gains in defect control and film properties.

It remains unknown how high the electronic transport mobility of $BaZrS_3$ can be, and what range of carrier type and concentration can be achieved in thin films. We find here that room temperature mobility is phonon-limited, even for highly-doped films. This is consistent with two of our prior observations, of very high dielectric constant, and rather low optical phonon frequency.[20,28] We are not aware of a framework to calculate the mobility of highly-polarizable semiconductors in the degenerate doping limit, this may be an opportunity for theory development. Of course, for device development, it will be important to reduce the carrier concentration and ideally to make both *n*- and *p*-type films. The successful demonstration of ambipolar doping in $SrHfS_3$ powder samples is encouraging for achieving similar in $BaZrS_3$ thin films.[23] Our results here suggest that sulfur vacancy concentration must be first minimized before aliovalent doping strategies are likely to succeed.

## Methods

We grow films on LaAlO$_3$ substrates by gas-source molecular beam epitaxy (MBE) at nominal substrate temperature of 900–1000°C following our previously reported growth procedures[13]. Resulting film thicknesses are typically between 20–40 nm depending on specific growth parameters. High-resolution x-ray diffraction (HRXRD) was measured using a Bruker D8 diffractometer configured with a Cu x-ray source and scintillation point-detector.

Exploratory sheet resistance measurements were done by linear four-point probe using a Signatone Pro4 test stand and Keithley 2601B source-measure unit. Samples with measurable conductivity undergo van der Pauw Hall measurements using a Lakeshore Cryotronics M91 FastHall measurement controller with a Quantum Design Physical Properties Measurement System (PPMS) for temperature and magnetic field control. Bidirectional sweeps of the magnetic field were performed for each measurement, with magnitudes of at least 2 T. Larger field values were used as-needed to ensure adequate signal-to-noise ratios. Ti (5 nm) / Au (200 nm) electrical contact pads were deposited on samples by sputtering and then contacted using a Van der Pauw press-contact assembly with spring-loaded gold pins (Wimbush Science and Technology). For sufficiently conductive samples, direct contact with the press-contact assembly provided suitable ohmic contacting behavior, so sputtered contacts were not needed.

Tube furnace annealing was performed in a vacuum-assisted quartz tube furnace with a 40 mm outer diameter tube. After loading each sample, three pump / purge cycles were performed with Ar purge gas (typical base pressures were 3 – 20 mTorr). H$_2$S flow was initiated at the selected flow rate before ramping to process temperature at approximately 10°C / minute. After holding for 30 minutes at the process temperature the furnace was allowed to cool naturally with the process gas flow maintained.

Ultra-high vacuum (UHV) H$_2$S plasma annealing was performed in our MBE growth chamber (10$^{-8}$ Torr base pressure). After about 15 weeks of air exposure, Sample A5 was reloaded into the UHV chamber. With the substrate unheated, an H$_2$S plasma was ignited in a SPECS RF source and ramped to a forward power of 600 W. While under plasma exposure, the substrate was heated to 200 °C, held for 30 minutes, then cooled naturally. Plasma source ramp-down began once the substrate temperature dropped below 65 °C.

Photoconductivity was measured using a Keithley 6517b electrometer under 1 V of applied bias. Illumination was provided by a Thorlabs SOLIS-1A 6500 K cold white LED driven at 5 A, resulting in an irradiance of 100 mW/cm$^2$ at the sample plane as measured by a Thorlabs S310C thermal power sensor.

SEM image acquisition was done in a Zeiss Gemini 450 field-emission SEM operating at 1.5 kV accelerating voltage using the Everhart Thornley secondary-electron detector.

ToF-SIMS measurements were carried out using TOF.SIMS$^5$-NSC instrument (ION.TOF Gmb), which combines mass spectrometer with atomic force microscope in the same vacuum chamber with vacuum levels 1 – 5×10$^{-8}$ mbar. Bi$_3^+$ liquid metal ion gun (30 keV energy, ~0.5 nA current in DC mode, 120 nm spot size) was used as a primary ion source to extract secondary ions from the surface of studied sample. This was complemented by Cs$^+$ sputter ion source (1 keV energy, 80 nA current and 10-20 μm spot size). Measurements were carried out in non-interlaced mode, where each scan by Bi$_3^+$ (100x100 μm$^2$, 2.4 s) was followed by sputtering cycle with Cs$^+$ (500×500 μm$^2$, 3.3 s). Atomic force microscopy was further used to measure depth of the resulting craters and further calibrate depth profiles measured by ToF-SIMS.

Steady-state photoluminescence spectroscopy (PL) was performed using a Horiba SMS spectroscopy system fitted with a10x infinity-corrected objective, iHR320 spectrometer, 600 gr/mm 750 nm blaze holographic grating, and Syncerity silicon CCD. BZS samples were excited with both 405 nm and 635 nm Horiba DeltaDiode lasers to probe surface and bulk behavior respectively. Measurements remove laser signal using appropriate longpass filters.

**Acknowledgements**

This research was supported by the Air Force Office of Scientific Research under grant no. FA9550-23-1-0695. This research was supported in part by the United States-Israel Binational Science Foundation (BSF) under grant no. 2020270. This research was supported in part by the National Science Foundation under grant no. DMR-2224948. This research was supported in part by the Sagol Weizmann-MIT Bridge Program. J.V.S. and J.D. acknowledge support from the National Science Foundation Graduate Research Fellowship, grant no. 1745302. We acknowledge support from AFOSR under award no. FA9550-23-1-0157. ToF-SIMS characterization was conducted at the Center for Nanophase Materials Sciences, which is a DOE Office of Science User Facility, and using instrumentation within ORNL's Materials Characterization Core provided by UT-Battelle, LLC under Contract No. DE-AC05-00OR22725 with the U.S. Department of Energy. This work was performed in part at MIT.nano.

**Supplemental Information**

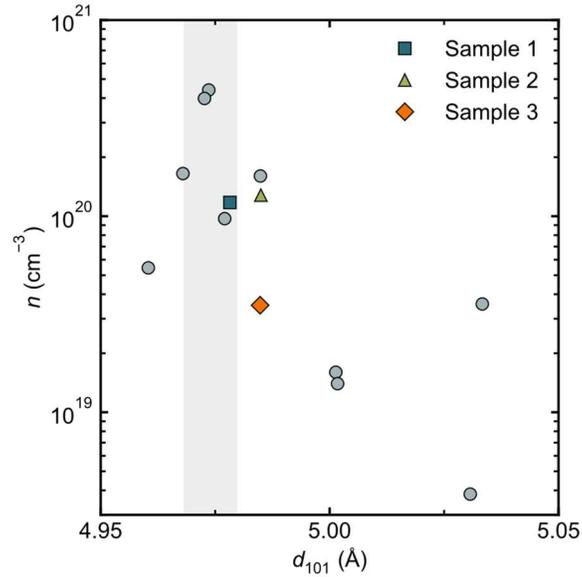

**Figure S1:** Hall carrier concentration measured at 300 K as a function of HRXRD-measured $d_{101}$ for the set of samples shown in Figure 1a. The square is Sample 1, the triangle is Sample 2, and the rotated square is Sample 3. Circles indicate other samples for which further characterization is not included in this report.

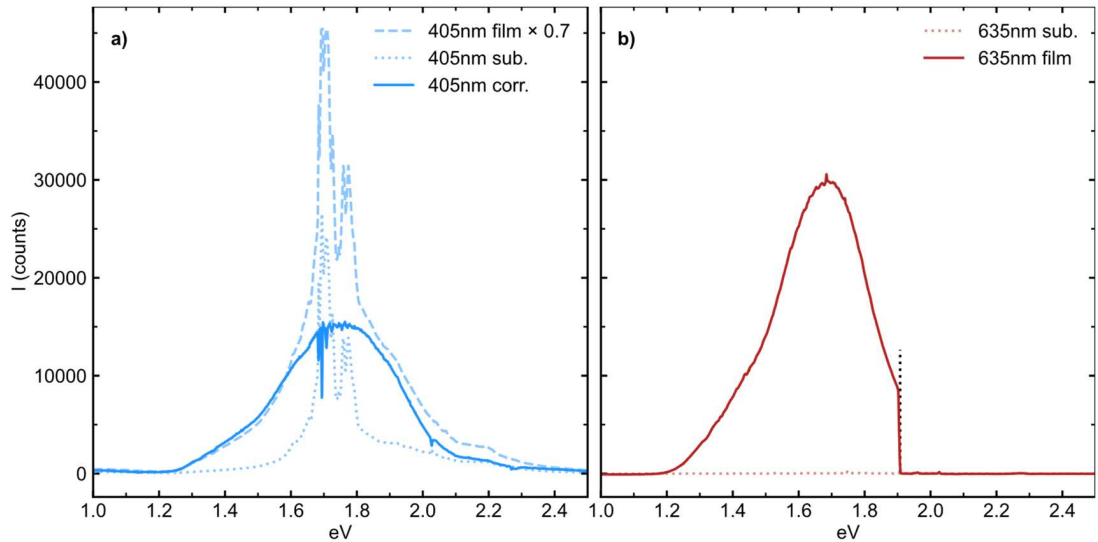

**Figure S2:** Photoluminescence (PL) data for Sample 1, measured at excitation wavelengths of 405 nm and 635 nm. **a)** Data for 405 nm excitation. Sharp emission features from the LaAlO$_3$ substrate appear in the spectra measured on the film (dashed line) and bare substrate (dotted line) which obscure the film emission. Emission we attribute to the film is shown by the solid line, which is obtained by scaling the bare substrate data by a factor of 2.15 and subtracting from the film data. **b)** Raw PL data for 635 nm excitation of the film (solid line) and bare substrate (dashed line). Substrate features were not observed for this excitation wavelength. Vertical line indicates position of the laser line edge filter.

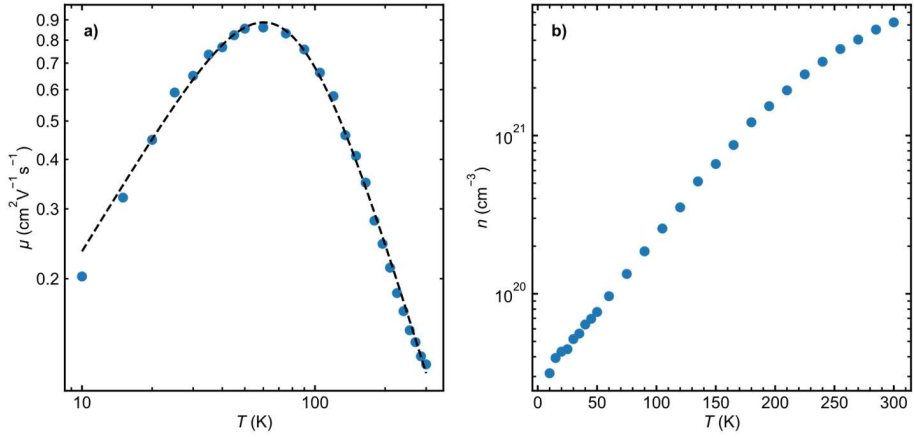

**Figure S3:** Temperature-dependent Hall measurement results for Sample P1, a polycrystalline BZS film grown on an $Al_2O_3$ substrate. **a)** Temperature-dependent Hall mobility, with the dashed curve showing a least-squares fit to the expression: $\mu^{-1} = AT^\alpha + BT^\beta$, with best-fit parameters of $\alpha = -1.9$ and $\beta = 1.0$. **b)** Temperature-dependent electron density.

**Table S1:** Annealing conditions and sample properties for annealed Samples A1–A5 and Sample 3. "-" indicates values that were immeasurable or were not characterized for this work. $\Delta n$ is the annealing-induced reduction in carrier concentration reported in **Fig. 4b**. Thicknesses were determined from SIMS profiles. Equipment issues caused the $H_2S$ flow rate for Sample A4 to drop from 20 sccm to 10 sccm during film processing.

| Sample | A1 | A2 | A3 | A4 | A5 | 3 |
|---|---|---|---|---|---|---|
| Furnace | Tube | Tube | Tube | Tube | UHV | $PH_3$ reactor |
| Gas / flow (sccm) | (vacuum) / 0 | $H_2S$ / 2 | $H_2S$ / 10 | $H_2S$ / 20-10 | $H_2S$ / plasma | $PH_3$ / 30 |
| Temperature (°C) | 150 | 150 | 500 | 500 | 200 | 400 |
| Time (min) | 30 | 30 | 30 | 30 | 30 | 30 |
| $n$, pre-anneal (cm$^{-3}$) | 2.20E+19 | 1.17E+21 | 3.24E+19 | 1.55E+20 | 6.94E+20 | - |
| $n$, post-anneal (cm$^{-3}$) | - | 1.85E+19 | - | - | 6.67E+19 | - |
| $\Delta n$ (cm$^{-3}$) | 2.20E+19 | 9.83E+19 | 3.24E+19 | 6.94E+20 | 6.28E+20 | - |
| Thickness (nm) | 45 | 25 | 33 | 32 | 36 | - |
| Morphology | closed | partially open | closed | partially open | partially open | closed |

**Table S2:** Sample names cross-referenced to our internal database names.

| Sample Name Here | Sample Name in Database |
|---|---|
| Sample 1 | G156 |
| Sample 2 | G172 |
| Sample 3 | G162 |
| Sample A1 | G154 |
| Sample A2 | G163d |
| Sample A3 | G164c |
| Sample A4 | G163a |
| Sample A5 | G157 |
| Sample P1 | G086 |
| Sample N | G173 |

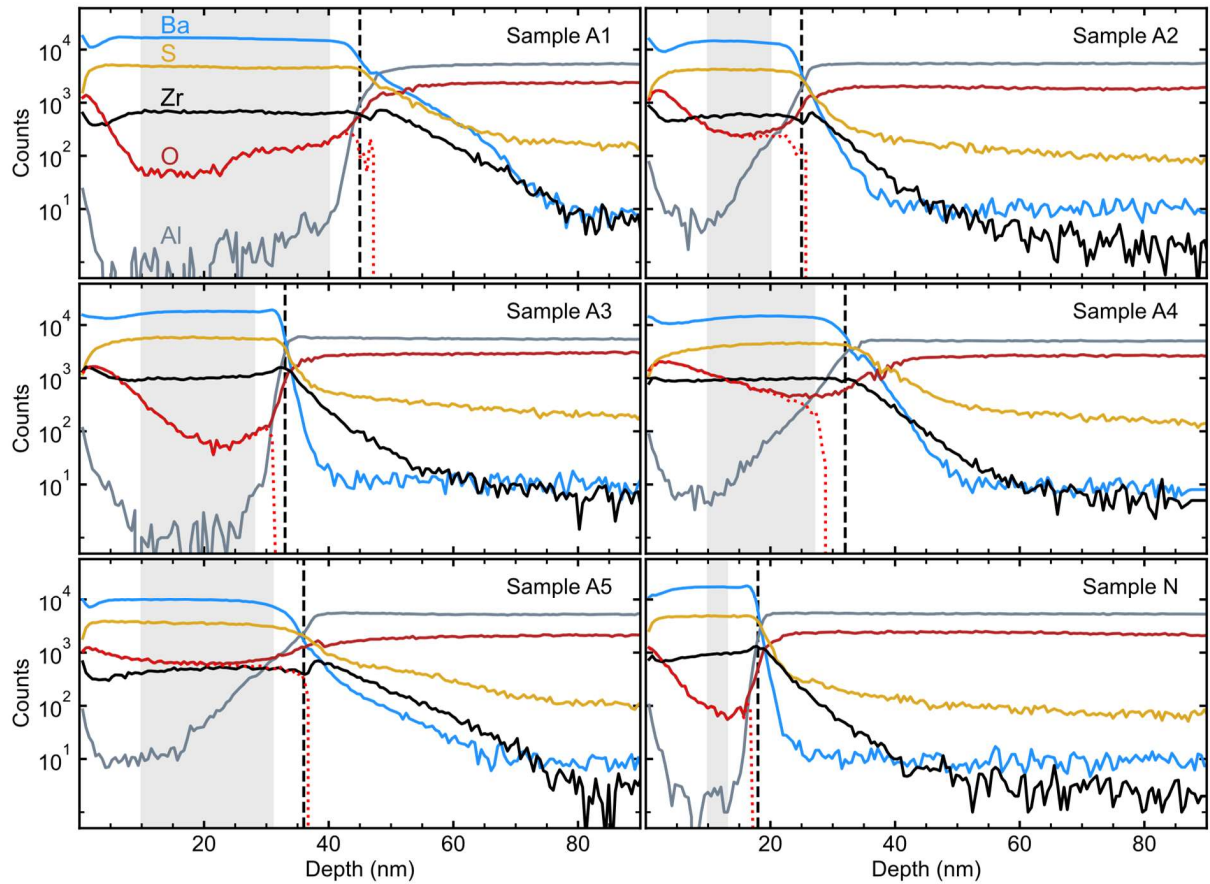

**Figure S4:** Depth-resolved SIMS compositional line scans showing signal intensities for Ba, S, Zr, O, and Al ions for annealed Samples A1–A5 and the as-grown Sample N. In each panel, the vertical dashed line shows the film thickness, and signal averaging over the grey shaded region was used in the analysis shown in Fig 4b. The solid red lines show total O ion intensities, while the broken red lines show the intensity attributed to the film (see Supplemental Note 1).

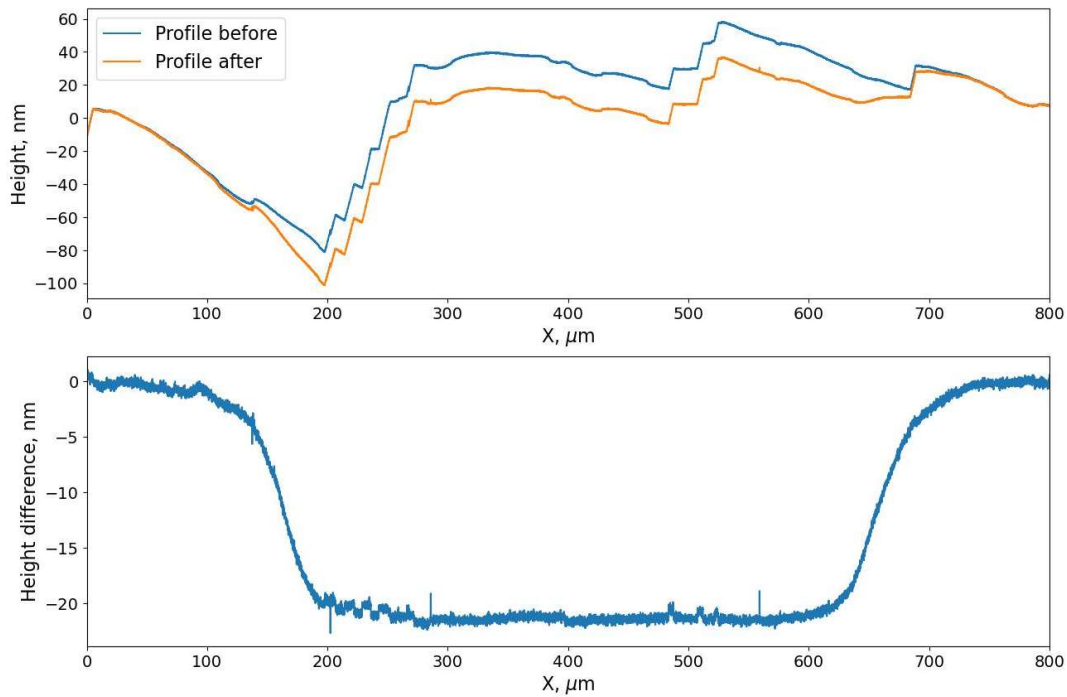

**Figure S5:** Atomic-force microscopy (AFM) line scans used for calibration of SIMS depth profile on Sample N. Raw scans (bottom panel) are shown before and after ion-beam milling for ~160 s. Lateral variations in height are due to twin defects in the $LaAlO_3$ substrate. The differential scan (top panel) clearly shows the etched area and allows to accurately identify etching rate.

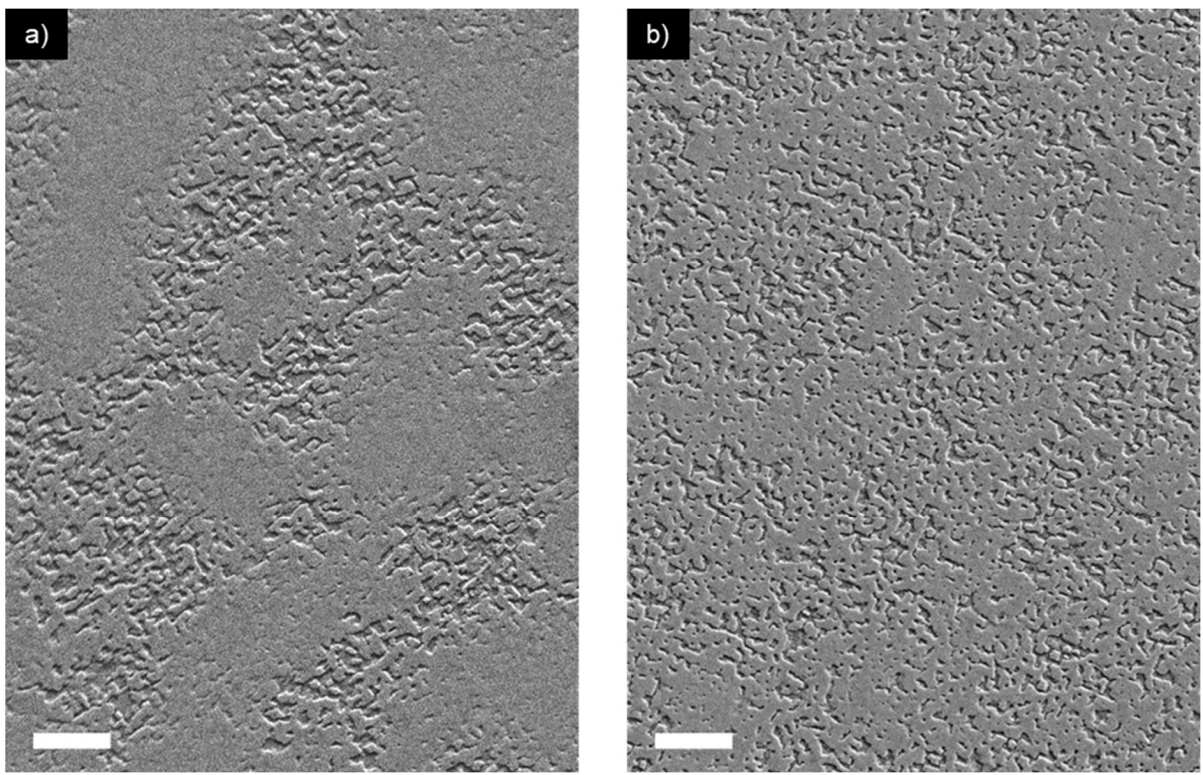

**Figure S6:** SEM micrographs of Sample A4 as grown (a) and after annealing (b). No reference fiducials were used, so micrographs do not show the exact same region of the sample. Scalebars are 1 μm, with each panel showing a 10 x 7.5 μm field of view.

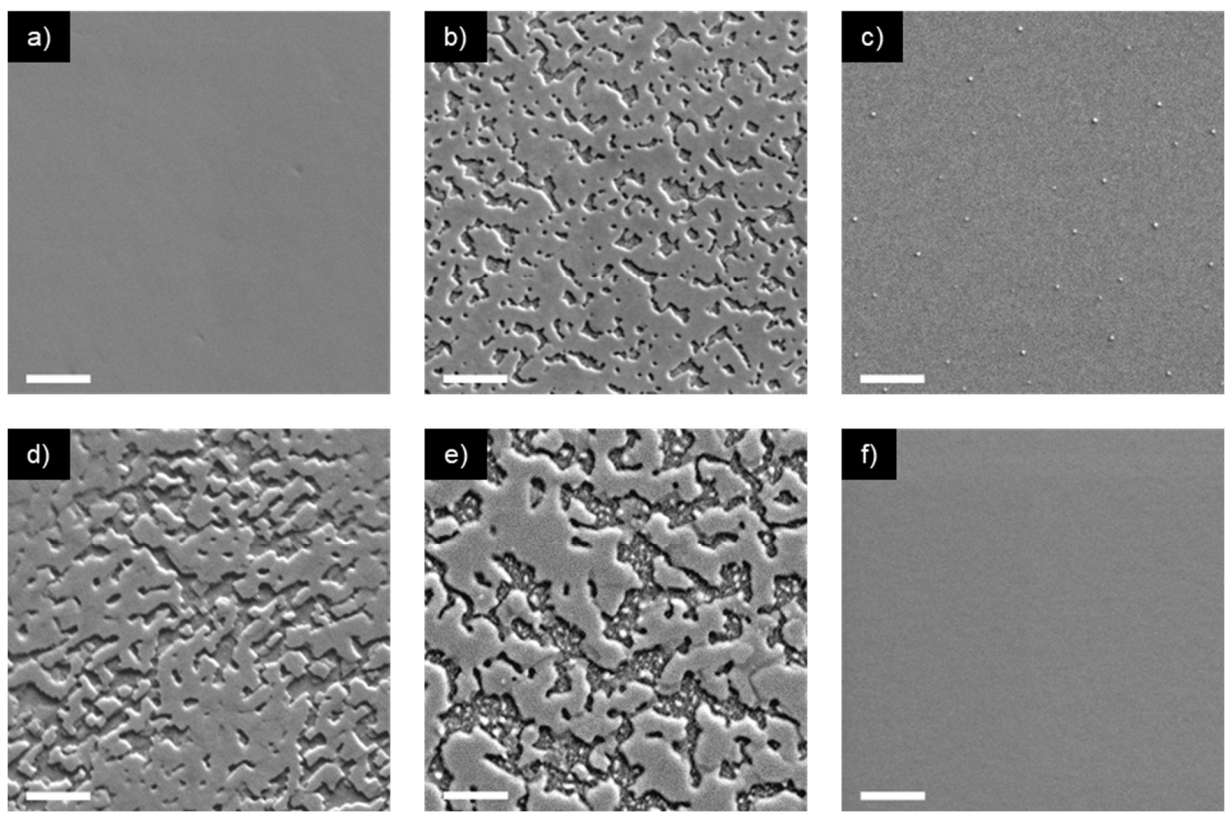

**Figure S7:** SEM micrographs of as-grown films used for SIMS study. a) Sample A1, b) Sample A2, c) Sample A3, d) Sample A4, e) Sample A5, f) Sample N, the reference sample used in SIMS study. All scalebars are 500nm, and each panel shows a 3 x 3 μm field of view.

**Supplementary Note 1**

Several of the films used in the SIMS study had incomplete surface coverage (**Fig. S7**), which complicates analysis of the ToF-SIMS oxygen signal. The films are grown on oxide substrates, so apparent oxygen signal observed within the depth of the film could either be from impurities in the film or from exposed regions of the substrate. This was taken into account by assuming a fixed substrate stoichiometry and therefore a constant ratio between the substrate's Al signal and the substrate's O signal. Through this treatment, the observed total O signal can be separated into film and substrate contributions. The resulting O line scans attributed to the film are shown by the dashed lines in **Fig. S4**, and are found to deviate insignificantly from the total O signal over the regions used in the averaging analysis.